\documentclass[amsmath,amssymb,prb,aps,showpacs,superscriptaddress,preprint]{revtex4-1}
\usepackage{amsmath,amsfonts,amssymb,amsthm,graphics,graphicx,epsfig,bbm}
\usepackage[colorlinks=true,citecolor=blue,linkcolor=blue,urlcolor=blue]{hyperref}
\usepackage[usenames]{color}
\usepackage{graphicx}
\usepackage{subfigure}
\usepackage{amsmath}
\usepackage{epsfig,times}
\usepackage{dcolumn}
\usepackage{bm}
\usepackage{color}
\usepackage{times}
\usepackage{epstopdf}
\usepackage{amssymb}
\usepackage{amstext}
\usepackage{latexsym}
\usepackage{hyperref}
\usepackage{amsfonts}
\usepackage{psfrag}
\usepackage{xcolor}
\usepackage[normalem]{ulem}
\usepackage{perpage}
\MakePerPage{footnote}

\def\beq{\begin{equation}}
\def\eeq{\end{equation}}
\def\beqa{\begin{eqnarray}}
\def\eeqa{\end{eqnarray}}

\newcommand{\ket}[1]{\vert #1 \rangle}

\newcommand\RS{\bgroup\markoverwith
{\textcolor{red}{\rule[0.5ex]{2pt}{1.0pt}}}\ULon}

\begin{document}

\title{Scalable quantum memory in the ultrastrong coupling regime}

\author{T. H. Kyaw\footnote{Corresponding author: thihakyaw@nus.edu.sg}}
\affiliation{Centre for Quantum Technologies, National University of Singapore, 3 Science Drive 2, Singapore 117543, Singapore}
\author{S. Felicetti}
\affiliation{Department of Physical Chemistry, University of the Basque Country UPV/EHU, Apartado 644, E-48080 Bilbao, Spain}
\author{G. Romero}
\affiliation{Department of Physical Chemistry, University of the Basque Country UPV/EHU, Apartado 644, E-48080 Bilbao, Spain}
\author{E. Solano}
\affiliation{Department of Physical Chemistry, University of the Basque Country UPV/EHU, Apartado 644, E-48080 Bilbao, Spain}
\affiliation{IKERBASQUE, Basque Foundation for Science, Maria Diaz de Haro 3, 48013 Bilbao, Spain}
\author{L.-C. Kwek}
\affiliation{Centre for Quantum Technologies, National University of Singapore, 3 Science Drive 2, Singapore 117543, Singapore}
\affiliation{Institute of Advanced Studies, Nanyang Technological University, 60 Nanyang View, Singapore 639673, Singapore}
\affiliation{National Institute of Education, Nanyang Technological University, 1 Nanyang Walk, Singapore 637616, Singapore}

\date{\today}

\begin{abstract}
Circuit quantum electrodynamics, consisting of superconducting artificial atoms coupled to on-chip resonators, represents a prime candidate to implement the scalable quantum computing architecture because of the presence of good tunability and controllability. Furthermore, recent advances have pushed the technology towards the ultrastrong coupling regime of light-matter interaction, where the qubit-resonator coupling strength reaches a considerable fraction of the resonator frequency. Here, we propose a qubit-resonator system operating in that regime, as a quantum memory device and study the storage and retrieval of quantum information in and from the $Z_2$ parity-protected quantum memory, within experimentally feasible schemes. We are also convinced that our proposal might pave a way to realize a scalable quantum random-access memory due to its fast storage and readout performances.
\end{abstract}

\maketitle

Analogous to the classical computer processors, a quantum processor inevitably requires memory cell elements \cite{Pritchett2005, Morimae2010, Reim2011, Saito2013} to store arbitrary quantum states in efficient and faithful manner. In particular, these memory devices might be needed and useful for storing and retrieving qubits in a fast timescale between the quantum processor and the memory elements, similar to a classical random-access memory \cite{RAM}. That would require ability to store information for a short time with fast storage and readout responses. Here, we propose a quantum memory implemented on circuit quantum electrodynamics (cQED)~\cite{Blais04,Chiorescu04,Wallraff04} with fast storage and retrieval responses. The memory cell operates in the ultrastrong coupling (USC) regime of light-matter interaction~\cite{Bourassa2009,Niemczyk2010,Pol2010}, where the qubit-resonator coupling strength approaches a significant fraction of the resonator frequency. In addition, a key ingredient towards realizing our scheme is the existense of the $Z_2$ parity symmetry \cite{Casanova2010,Braak2011}, which allows us to encode quantum information in partiy-protected states that are robust against certain environmental noises \cite{Doucot2005,Nataf2011}. 

In this report, we propose a scalable quantum memory cell element based on the cQED architecture, comprising of a superconducting flux qubit galvanically coupled \cite{Bourassa2009} to a microwave resonator. In particular, we study the storage and retrieval of single- and two-qubit states, while the input states are in the form of flying microwave photons \cite{Wenner2013, Houck2007, Yin2013, Srinivasan2014}. These  processes can be carried out with good fidelity even with the presence of noise. We are also convinced that our memory can be scaled up to store large number of qubits since the cQED architecture provides very high level of controllability \cite{Barends2014,Jeffrey2014} and scalability \cite{Chow2014}. In this way, we believe our proposal might pave a way towards scalable quantum random-access memory (QRAM)~\cite{Giovannetti2008_1,Giovannetti2008_2} and distributed quantum interconnects~\cite{Cirac1999,Kimble2008}, which in turn might steer towards novel applications ranging from entangled-state cryptography~\cite{Ekert1991,Bennett1995}, teleportation~\cite{Bennett1993}, purification~\cite{Bennett1996,Deutsch1996}, fault-tolerant quantum computation~\cite{Nigg2013} to quantum simulations.

\begin{figure}[h]
\centering
\includegraphics[width=0.6\textwidth]{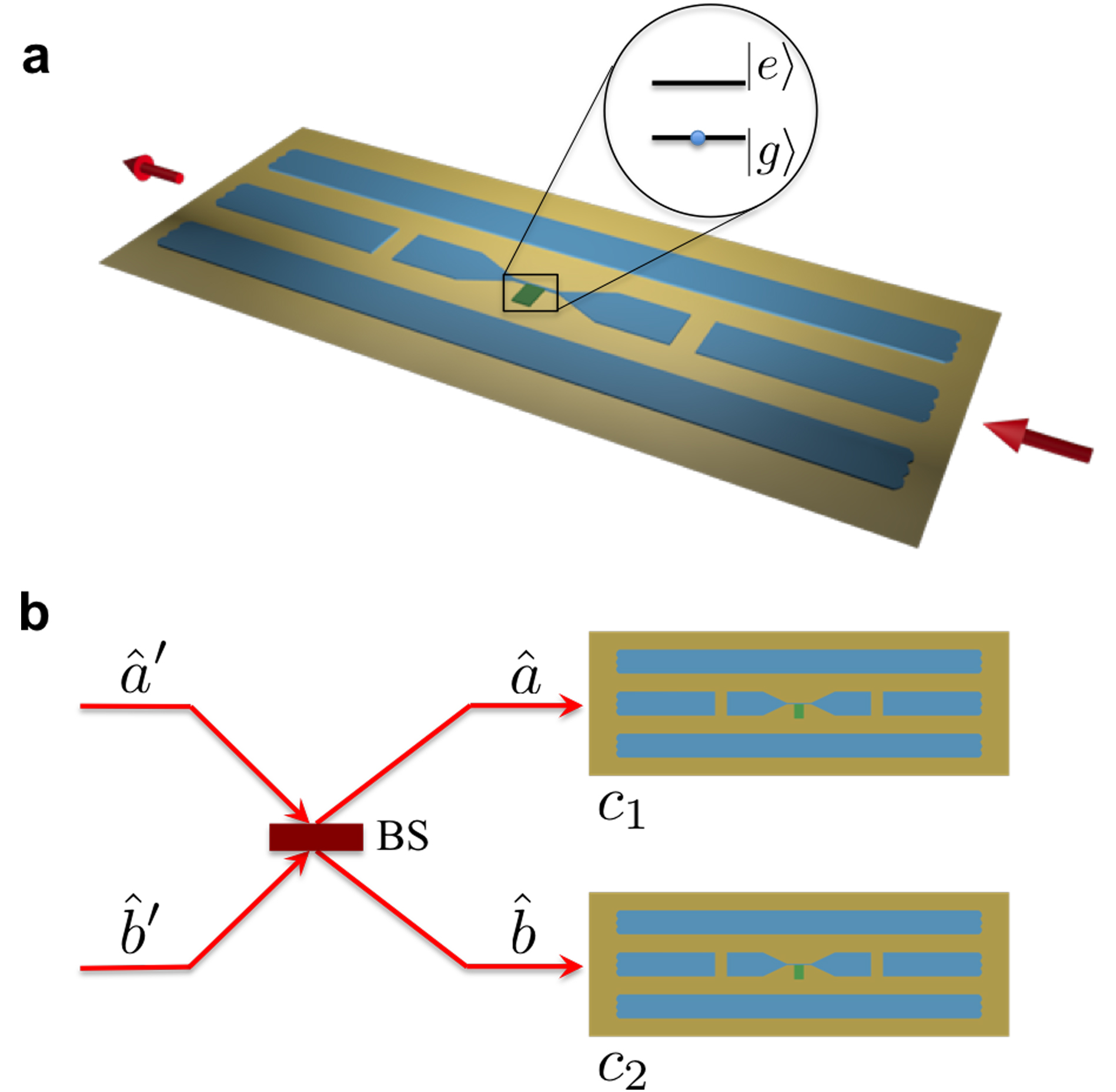}
\caption{}
\label{Fig1}
\end{figure}

The qubit-resonator system operating in the USC regime, as shown in Fig. \ref{Fig1}, exhibits a $Z_2$ parity symmetry and its dynamics is governed by the quantum Rabi Hamiltonian~\cite{Braak2011}
\begin{equation}
H_{\rm Rabi} = \frac{\hbar\omega_{eg}}{2} \sigma_z +\hbar \omega_{\rm cav} a^\dagger a + \hbar \Omega\sigma_x(a + a^\dagger),
\label{HRabi}
\end{equation}
where $\omega_{eg}$, $\omega_{\rm cav}$, and $\Omega$ stand for the qubit frequency, cavity frequency, and qubit-resonator coupling strength, respectively. In addition, $a(a^{\dagger})$ is the bosonic annihilation(creation) operator, and $\sigma_{x,z}$ are the Pauli matrices of the qubit. A compelling feature of Hamiltonian~(\ref{HRabi}) is that for ratios $\Omega/\omega_{\rm cav} \gtrsim~0.8$, the ground and first excited states can be approximated as
\beqa
\ket{\psi_{G}}&\simeq& \frac{1}{\sqrt{2}} (\ket{-\alpha}\ket{+} - \ket{\alpha}\ket{-}) , \nonumber \\
\ket{\psi_{E}}&\simeq& \frac{1}{\sqrt{2}} (\ket{-\alpha}\ket{+} + \ket{\alpha}\ket{-}),
\label{eigenstates}
\eeqa
where $\ket{\alpha}$ is a coherent state for the resonator field with amplitude $|\alpha|$ $= \Omega/\omega_{\rm cav}$, and $\ket{\pm}=(\ket{e}\pm \ket{g})/\sqrt{2}$ are the eigenstates of $\sigma_x$. The states $\ket{\psi_{G/E}}$ form a robust parity-protected qubit~\cite{Nataf2011} whose coherence time can be up to $\tau_{\rm coh}\gtrsim 10^5 /\omega_{eg}$. In the following, we outline a protocol that allows storage and retrieval of quantum information to and from this qubit. It is achieved by adiabatically tuning the qubit-resonator coupling strength, from the Jaynes-Cummings (JC) to USC regime. In particular, we propose a USC memory cell element (see Fig.~\ref{Fig1}a) that can be designed by the flux-qubit architecture presented in Ref.~\citenum{Romero2012}, which provides a tunable qubit-resonator coupling (see Supplementary information). The latter can be implemented by using a superconducting quantum interference device (SQUID) as proposed for qubit-qubit coupling in Refs.~\citenum{Makhlin1999, Grajcar2006}.
\\ \\
{\bf \large Results}\\
\textbf{Generating and catching flying qubits}. Photons propagating through linear devices are well-suited as information carriers because they possess long coherence length and can be encoded with useful information. In our case, a flying microwave photon is generated from cQED platforms \cite{Wenner2013, Houck2007, Yin2013, Srinivasan2014}, and a qubit is encoded in a linear superposition of zero $(\ket{0_F})$ and one photon $(\ket{1_F})$ Fock states. Recently, it has been pointed out in Ref. \citenum{Srinivasan2014} that if a photonic wave packet emitted from a source has a temporally symmetric profile, it overcomes the impedance mismatch problem when a flying qubit impinges onto a resonator. With all these latest advancement in cQED technologies, we envision our memory cell be located on the pathway of a single microwave photon to accomplish quantum information storage (see Fig. \ref{Fig1}).\\

\begin{figure}[h]
\centering
\includegraphics[width=1.1\textwidth]{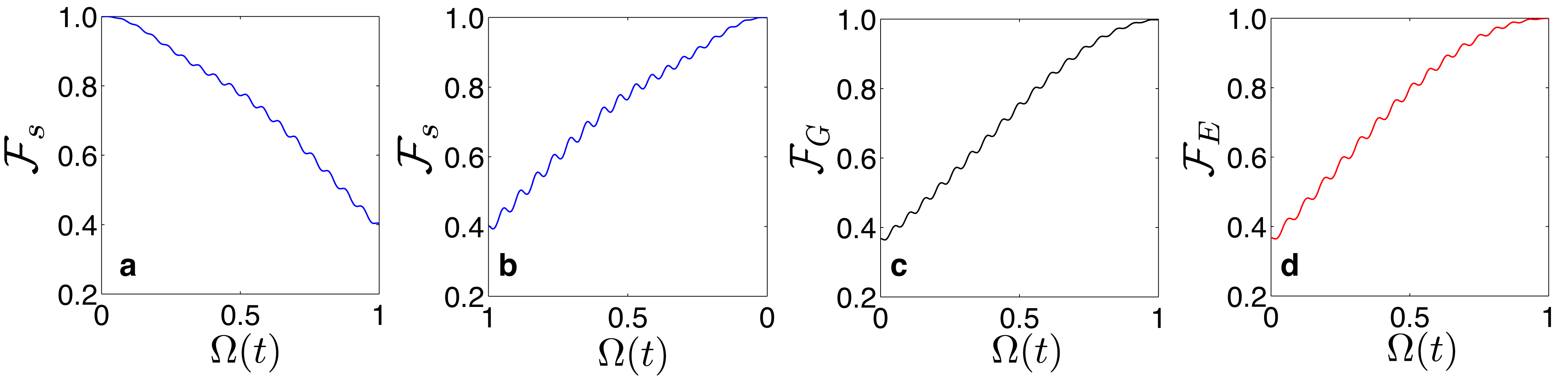}
\caption{}
\label{Fig2}
\end{figure}

\noindent\textbf{Storage and retrieval processes}. The storage of quantum information into our USC memory cell is realized within three steps. At first, we cool down the system to reach its ground state in the USC regime. Secondly, the qubit frequency is tuned to be off-resonant with the resonator frequency, i.e., $\omega_{\rm{cav}}>\omega_{eg}$, while the qubit-resonator coupling stregth $\Omega$ is adiabatically tuned towards the strong coupling regime where $\Omega/\omega_{\rm cav}\ll 1$, where the coupling is much larger than any decoherence rate in the system. In this regime, the ground and first excited states of the qubit-resonator system are $\ket{\psi_0}=\ket{g}\otimes\ket{0}$ and $\ket{\psi_1}=\ket{e}\otimes\ket{0}$, respectively. Here, the states $\ket{g}$ and $\ket{e}$ stand for the ground and excited states of the qubit, while $\ket{0}$ stands for the vacuum state of the resonator. Since we have adiabatically tuned the coupling from the USC to the strong coupling regime, our initial USC ground state is then mapped to the JC ground state, i.e., $\ket{\psi_0}=\ket{g}\otimes\ket{0}$. At this stage, our memory cell is ready for information storage. When a flying qubit with an unknown quantum state $\ket{\Psi_F} = \alpha_F\ket{0_F}+\beta_F\ket{1_F}$ comes in contact with the cell as shown in Fig. \ref{Fig1}a, the encoded information from the flying qubit is transferred to the flux qubit due to the JC dynamics. Therefore, the state of our system becomes $\ket{\psi_s} = (\alpha_F\ket{g}+\beta_F\ket{e})\otimes \ket{0}$. At last, we turn on the qubit-resonator coupling adiabatically towards the USC regime. For simplicity, we consider a linear adiabatic switching scheme such that $\Omega(t)=(\cos(f) - \Delta f\sin(f) t/T)\Omega_0$, with $T$ total evolution time and $f=\phi_{\text{ext}}/\phi_0$. Here, $\phi_{\text{ext}}$ is an external magnetic flux and $\phi_0=h/2e$ is the flux quantum (see Methods and Supplementary information for detailed definition). In Fig.~\ref{Fig2}a,b, we show the storage and retrieval processes for a quantum state $\ket{\psi_s} = \alpha_F\ket{\psi_0}+\beta_F\ket{\psi_1}$, and Fig.~\ref{Fig2}c,d show the ground state $\ket{\psi_0}$ and the first excited state $\ket{\psi_1}$ adiabatically follow the instantaneous eigenstates such that $\ket{\psi_0}\to\ket{\psi_G}$ and $\ket{\psi_1}\to\ket{\psi_E}$. In this manner, we can encode important information onto the parity-protected qubit basis. Retrieval (decoding) process is reverse of the storage process and is achieved by adiabatically switching off the qubit-resonator coupling strength from the USC to SC regime.

We note that the time for storage and retrieval of quantum information is several order of magnitude faster than the coherence time of the parity-protected qubit, which is about $T_{\rm coh}\sim40~\!\mu$s for a coupling strength $\Omega_0/\omega_{eg}\sim 1.5$~\cite{Nataf2011}. For instance, if we consider a flux qubit with energy $\omega_{eg}/2\pi\sim 2~{\rm GHz}$, and a cavity of frequency $\omega_{\rm cav}/2\pi \sim 5~{\rm GHz}$, our system reaches the USC regime with $\Omega_0/\omega_{\rm cav} = 0.6$. For the linear adiabatic switching scheme with the above parameters, we estimate total time for storage/retrieval of a qubit is about $\tilde{T}\approx 2-8\!$ ns.

At the end of an adiabatic evolution, the state $\ket{\tilde{\psi}}=\alpha_F\ket{\psi_G}+\beta_F\ket{\psi_E}$ is desired. However, the state after the evolution might become
$
\ket{\tilde{\psi}(T)} = \alpha_F \ket{\psi_G(T)}  +\beta_F e^{i\theta(T)} \ket{\psi_E(T)},
$
with a relative phase $\theta(T)$ resulting from the dynamical and geometrical effects~\cite{Berry_1984}. Hence, we need to keep track of a relative phase during the storage and retrieval processes. 

\begin{figure}[h]
\centering
\includegraphics[width=0.7\textwidth]{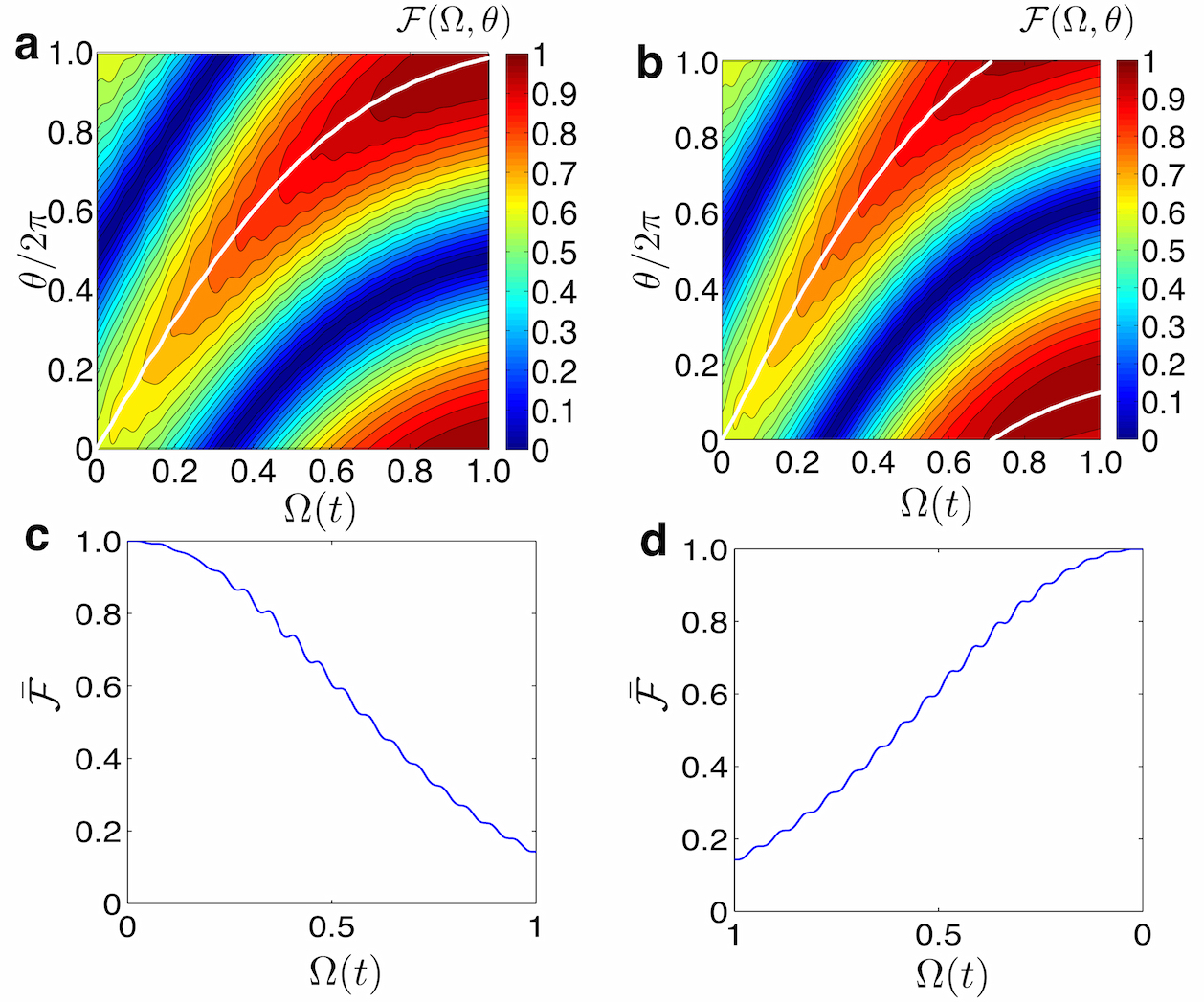}
\caption{}
\label{phase_contour}
\end{figure}

In order to find out which phase $\theta(t)$ optimizes the processes, in Fig.~\ref{phase_contour}a,b, we plot the fidelity $\mathcal{F}(\Omega,\theta)=|\langle \tilde{\psi}|\psi(t) \rangle|^2$ between the state $\ket{\tilde{\psi}} = \alpha_F \ket{\psi_G}  +\beta_F e^{i\theta} \ket{\psi_E}$ and the state $\ket{\psi(t)}$, which has adiabatically evolved from the initial state $\ket{\psi_s} = \alpha_F\ket{\psi_0}+\beta_F\ket{\psi_1}$. In these simulations, we find the fidelity over the landscape of $\theta\in [0,2\pi]$ versus the qubit-resonator coupling strength $\Omega(t)$, for two different total evolution time $T=105/\omega_{\rm cav}$ (Fig.~\ref{phase_contour}a), and $T=120/\omega_{\rm cav}$ (Fig.~\ref{phase_contour}b). White lines show the phase $\theta_{\rm opt}$, which optimizes the fidelity $\mathcal{F}$ for both cases. Notice that the maximum fidelity and the optimal phase $\theta$ depend strongly on the system parameters and the total evolution time $T$. Thus, we require, for each USC memory cell, to find out the parameter $T$ that maximizes the fidelity only once. When $T$ is known, the cell can always be operated at that specific parameter for storing and retrieving unknown quantum states. Therefore, the time $T$ might be a benchmark to characterize our potential USC quantum memory devices, in the same way as hard disk drives of the classical computer are being characterized by their seek time and latency.

\begin{figure}[h]
\centering
\includegraphics[width=1\textwidth]{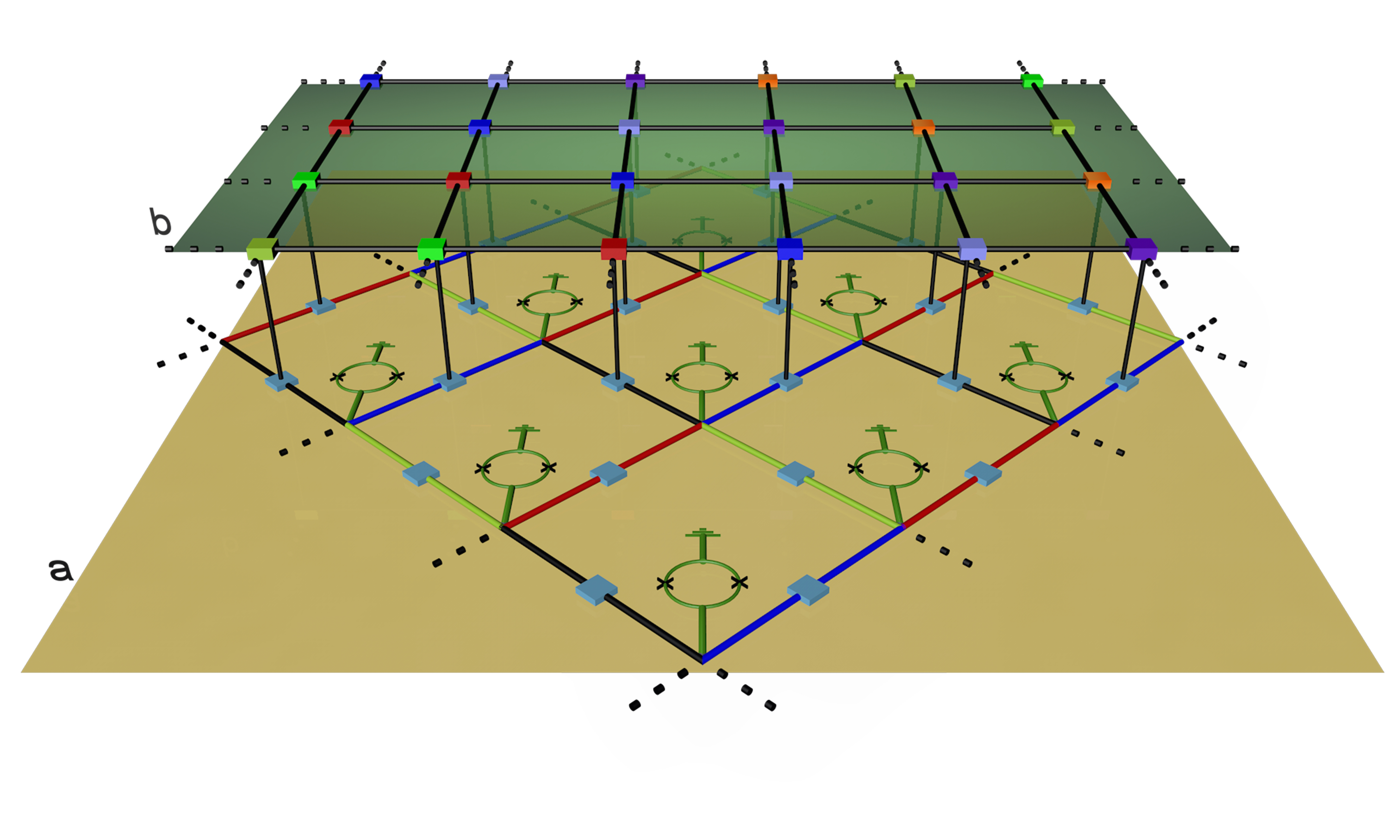}
\caption{}
\label{network}
\end{figure}

Additionally, storage and retrieval of entangled states in two separate USC cells is feasible. To demonstrate such a process, we let two bosonic fields to interact via the SQUID, simulating a Hong-Ou-Mandel setup~\cite{Wallraff2013} as shown in Fig.~\ref{Fig1}b. Let us suppose that we have an initial state $\ket{\psi_0}=\ket{0}_{\hat{a}'}\ket{1}_{\hat{b}'}$. After experiencing a beam splitter interaction, we have two-photon entangled state $\ket{\psi_0 '}=\frac{1}{\sqrt{2}}(\ket{0}_{\hat{a}}\ket{1}_{\hat{b}} + \ket{1}_{\hat{a}}\ket{0}_{\hat{b}})$, which enters two cavities $c_1$ and $c_2$, each containing a flux qubit prepared in its ground state. This process allows the cavities to be prepared in the state $\ket{\overline{\Psi}_0}=\frac{1}{\sqrt{2}}(\ket{0}_{c_1}\ket{1}_{c_2} + \ket{1}_{c_1}\ket{0}_{c_2})\otimes \ket{gg}$. Following the same procedure, we tune the qubits towards resonance with its respective cavity such that we arrive at the state $\ket{\Psi_0}=\frac{1}{\sqrt{2}}(\ket{ge} + \ket{eg})\otimes \ket{00}_{c_1 c_2}$. With our protocol, the state is eventually mapped to a parity-protected state $\ket{\tilde{\Psi}_0}=\frac{1}{\sqrt{2}}(\ket{\psi_G} \ket{\psi_E}  + \ket{\psi_E} \ket{\psi_G})$. In Fig.~\ref{phase_contour}c,d, we show the numerical simulations for the storage and retrieval processes of the entangled state $\ket{\Psi_0}$.
\\ \\
\textbf{{\large Discussion}}\\
We have presented the basic tools for building a quantum memory based on a cQED architecture that operates in the USC regime of light-matter interaction. The storage/retrieval process for unknown quantum states, be single-qubit or two-qubit entangled states, can be accomplished by adiabatically switching on/off the qubit-resonator coupling strength. 
As a scope, we propose the large-scale quantum memory network shown in Fig.~\ref{network}(a), where each edge of the memory network is constituted with our memory cell element. This architecture can pave the way for the implementation of a scalable QRAM, which might benefit from the fast storage and readout performances of superconducting circuits. In addition, each node in the network is connected to a SQUID that allows to selectively switch on/off interaction between neighbouring microwave cavities~\cite{Felicetti2014,Romero2012}, in order to implement quantum state transfer processes within the same layer (see Supplementary information). Ultimately, we would like to achieve a multilayer circuit architecture, where a quantum processor layer \cite{Marquardt2009} interfaces with the proposed memory layer as shown in Fig.\ref{network}. 
\\ \\
{\bf \large Methods}\\
{\bf Switchable quit-resonator coupling strength}. In the cQED architecture composed of a flux qubit galvanically coupled to an inhomogeneous resonator, the Hamiltonian that describes the dynamics reads 
\begin{equation}
	H = \frac{\hbar\omega_{eg}}{2} \sigma_z +\hbar \omega_{\rm cav} a^\dagger a + H_{\rm int},\label{Hamil_complete}
\end{equation}
with an effective tunable interaction Hamiltonian 
\begin{equation}
H_{\rm int} = -2E_J\beta\cos\Big(\pi \frac{\phi_{\rm ext}}{\phi_0}\Big) \sum_{n=1,2}(\Delta\psi)^n\sum_{\mu=x,y,z}c^{(n)}_{\mu}\sigma_{\mu},
\label{Hint}
\end{equation}
where $E_J$ is the Josephson energy, $\beta$ is a parameter that depends on the Josephson junctions size, $\phi_0 =h/2e$ is the flux quantum, and $\phi_{\rm ext}$ is an external flux through a superconducting loop. The latter in turn allows to switch on/off the qubit-resonator coupling strength. $\Delta \psi$ stands for the phase slip shared by the resonator and the flux qubit. And, the coefficients $c^{(n)}_\mu$'s can be tuned \cite{Romero2012,Makhlin1999,Grajcar2006} at will via additional external fluxes (see Supplementary information).\\ 

\noindent {\bf Adiabatic evolution}. We obtain the effective system Hamiltonian $H= \frac{\hbar\omega_{eg}}{2} \sigma_z ^j +\hbar \omega_{\rm cav} a^\dagger a + (\cos(f) - \Delta f\sin(f) t/T)\Omega_0\sigma_x ^j (a^\dagger +a)$ from Eqs.~(\ref{Hamil_complete}) and~(\ref{Hint}), if we consider an external flux that varies linearly with time according to $\phi_{\rm ext}=\bar{\phi}_0+(\Delta\phi) t/T$, where $\bar{\phi}_0$ is an offset flux and $\Delta\phi$ is a small flux amplitude. 

We remark that all our simulations presented so far assume no loss in both the qubit and resonator. Nonetheless, the open system analysis of a USC system \cite{Rid_2012} can be carried out by studying dynamics of the microscopic master equation (see Supplementary information). In Fig. \ref{Fig5}, we show numerical results for the storage and retrieval processes of an arbitrary superposed state $\ket{\psi_s}$ in presence of external noises. With our scheme and a simple decoherence model, we estimate fidelity of $\mathcal{F}_s =0.9939$ at the end of the retrieval process.

\vspace{0.2 cm}
{\setlength{\parindent}{0pt}
{\bf Acknowledgments}\\
The authors acknowledge support from the National Research Foundation \& Ministry of Education, Singapore; Spanish MINECO FIS2012-36673-C03-02; UPV/EHU UFI 11/55; Basque Government IT472-10; and CCQED, PROMISCE, SCALEQIT EU projects. We also thank Lorenzo Maccone and Seth Lloyd for their helpful comments.

\vspace{0.2 cm}
{\bf Author contributions}\\
All authors T.H.K., S.F., G.R., E.S. and L.-C.K. contributed equally to the results.

\vspace{0.2 cm}
{\bf Additional information}\\
\textbf{Supplementary information} accompanies this paper at 

\textbf{Competing financial interests:} The authors declare no competing financial interests.
}

\newpage
\noindent \textbf{Figure 1. Schematic of circuit-QED design for storage and retrieval of an unknown single- and two-qubit states.} (a) A USC memory cell element, composed of a qubit-resonator system operating at the USC regime. (b) Two flying microwave photons, with modes $\hat{a}'$ and $\hat{b}'$, come in and pass through a beam splitter (BS) implemented by a superconducting quantum interference device (SQUID) to form a two-qubit entangled state, which is then stored in two USC qubits located at a distance apart.\\

\noindent \textbf{Figure 2. Fidelity plots.} (a) Storage and (b) retrieval processes for a quantum state $\ket{\psi_s} = \alpha_F\ket{\psi_0}+\beta_F\ket{\psi_1}$. In both cases, we plot the fidelity between the initial $\ket{\psi}_s$ and the instantaneous state $\ket{\psi(t)}$, i.e., $\mathcal{F}_s=|\langle \psi_s|\psi(t)\rangle|^2$. Any arbitrary state $\ket{\psi} = u\ket{\psi_0}+v\ket{\psi_1}$ can be stored and retrieved with unit fidelity. (c) Fidelity between the approximated ground state in Eq.~(\ref{eigenstates}) and the instantaneous ground state $\mathcal{F}_G=|\langle \psi_G|\psi_G(t)\rangle|^2$. (d) Fidelity between the approximated first excited state in Eq.~(\ref{eigenstates}) and the instantaneous first excited state $\mathcal{F}_E=|\langle \psi_E|\psi_E(t)\rangle|^2$. For all the simulations, we choose the system parameters as $\omega_{\rm cav}=1$, $\omega_{eg}=0.1~\omega_{\rm cav}$, $\Omega_0=\omega_{\rm cav}$, and the total evolution $T=105/\omega_{\rm cav}$.\\

\noindent \textbf{Figure 3. Contour plots} of the fidelity $\mathcal{F}=|\langle \tilde{\psi}|\psi(t)\rangle|^2$ between the state $\ket{\tilde{\psi}} = \alpha_F \ket{\psi_G}  +\beta_F e^{i\theta} \ket{\psi_E}$ and the state $\ket{\psi(t)}$, which has adiabatically evolved from the initial state $\ket{\psi_s} = \alpha_F\ket{\psi_0}+\beta_F\ket{\psi_1}$. (a) The total evolution time is set to $T=105/\omega_{\rm cav}$. (b) The evolution time is $T=120/\omega_{\rm cav}$. For the above cases, the black lines stand for the phase which maximized the fidelity $\mathcal{F}$. In these simulations, the parameters are $\omega_{\rm cav}=1$, $\omega_{eg} = 0.1~\omega_{\rm cav}$, and $\Omega_0=\omega_{\rm cav}$. (c) Storage process for an entangled state $\ket{\Psi_0}=\frac{1}{\sqrt{2}}(\ket{ge} + \ket{eg})\otimes \ket{00}_{c_1 c_2}$. (d) Retrieval process. In both cases, we plot the fidelity between the initial state $\ket{\Psi_0}$, and the instantaneous state $\ket{\psi(t)}$, $\mathcal{\bar{F}}=|\langle \Psi_0|\psi(t)\rangle|^2$. In the simulations (c) and (d), we have chosen $\omega_{\rm cav}=1$, $\omega_{eg}=0.1~\omega_{\rm cav}$, $\Omega_0=\omega_{\rm cav}$, and the evolution time $T=105/\omega_{\rm cav}$.\\

\newpage
\noindent \textbf{Figure 4. A scalable quantum network.} (a) The light-matter interface operating at the ultrastrong coupling regime may be envisioned as a set of microwave cavities connected, at the nodes, by SQUID devices that allow to switch on/off the cavity-cavity interaction. Notice that each cavity is represented with different colors (red, black, blue and green) that stand for different lengths to assure the manipulation of specific pairwise interactions (see Supplementary information). In addition, on each edge of the memory array, there is a memory cell made of a USC entity (blue square) to store an arbitrary quantum information in a specific location. (b) \textbf{Integrated quantum processor}. A 2D cavity grid with a qubit distribution (rectangular boxes) represented in various colors is shown here. It was previously shown in Ref. \citenum{Marquardt2009} that such a cavity grid may provide a scalable fault-tolerant quantum computing architecture with minimal swapping overhead. Data transfer between the two layers may be done via cavity buses (vertical black colored lines connecting layer a and b).\\

\noindent \textbf{Figure 5. Fidelity plots.} (a) Storage process for a quantum state $\ket{\psi_s} = \alpha_F\ket{\psi_0}+\beta_F\ket{\psi_1}$. (b) Retrieval process. In both cases, we plot the fidelity between the initial $\ket{\psi}_s$ and the instantaneous state $\ket{\psi(t)}$, i.e., $\mathcal{F}_s=|\langle \psi_s|\psi(t)\rangle|^2$, while we introduce external noises $\Gamma_x =\Gamma_y =\Gamma_z=10^{-3}\omega_{eg}$ and $\Gamma_r=10^{-4}\omega_{eg}$ into our close system with the help of the microscopic derivation \cite{Rid_2012}. Here, $\Gamma_x, \Gamma_y, \Gamma_z$ and $\Gamma_r$ are rates of bit-flip noises, dephasing noise of the qubit, and the resonator relaxation, respectively. The system parameters are $\omega_{\rm cav}=1$, $\omega_{eg}=0.1~\omega_{\rm cav}$, $\Omega_0=\omega_{\rm cav}$, and the total evolution time is $T=105/\omega_{\rm cav}$.

\begin{figure}[h]
\centering
\includegraphics[width=0.8\textwidth]{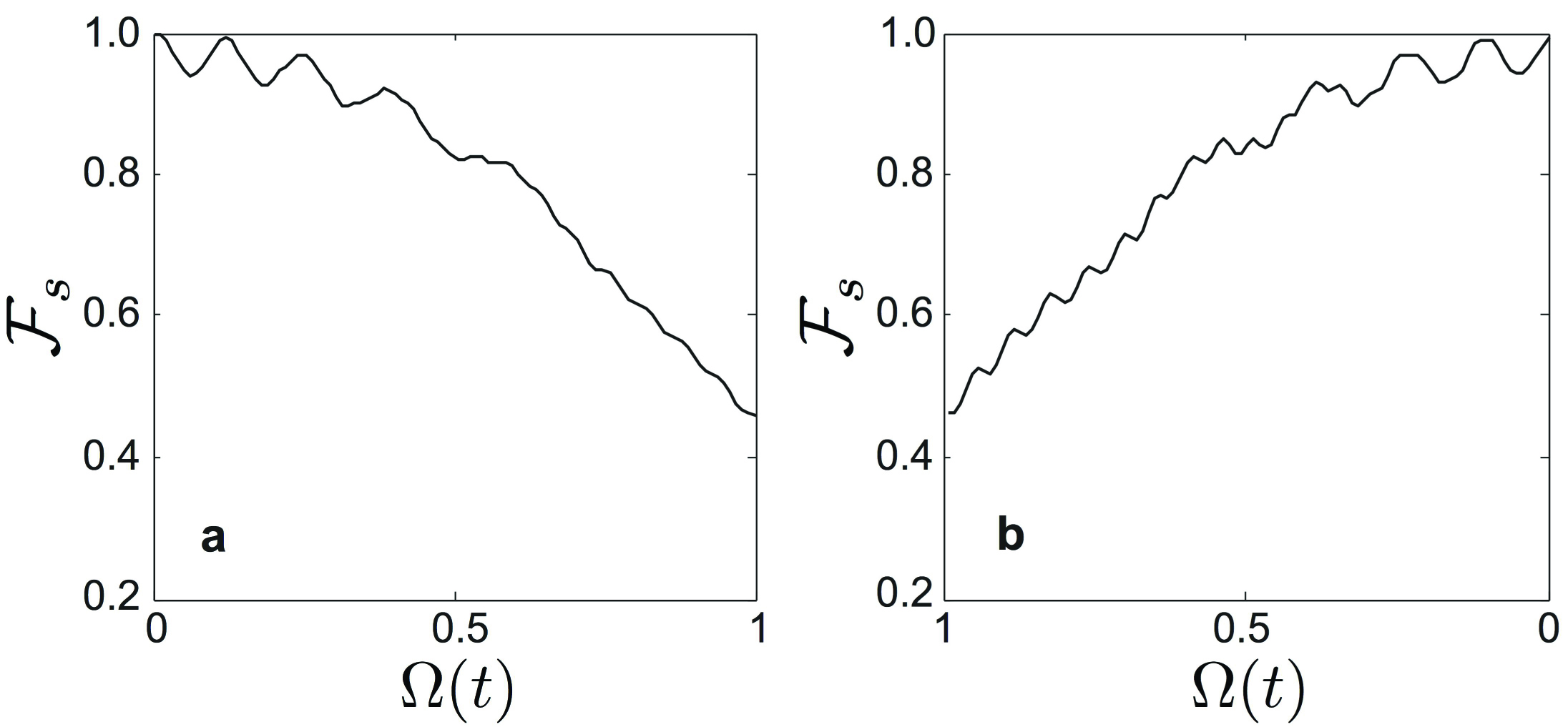}
\caption{}
\label{Fig5}
\end{figure}

\end{document}